\documentclass[12pt]{article}
\usepackage{cite,psfig}

\newcommand{\newc}{\newcommand}
\newc{\gsim}{\lower.7ex\hbox{$\;\stackrel{\textstyle>}{\sim}\;$}}
\newc{\lsim}{\lower.7ex\hbox{$\;\stackrel{\textstyle<}{\sim}\;$}}
\newc{\gev}{\,{\rm GeV}}
\newc{\mev}{\,{\rm MeV}}
\newc{\ev}{\,{\rm eV}}
\newc{\kev}{\,{\rm keV}}
\newc{\tev}{\,{\rm TeV}}

\newc{\mz}{M_Z}
\newc{\mpl}{M_*}
\newc{\mw}{m_{\rm weak}}
\def\beq{\begin{equation}}
\def\eeq{\end{equation}}
\def\bea{\begin{eqnarray}}
\def\eea{\end{eqnarray}}
\newc{\ie}{{\it i.e.}}          \newc{\etal}{{\it et al.}}
\newc{\eg}{{\it e.g.}}          \newc{\etc}{{\it etc.}}
\newc{\cf}{{\it c.f.}}
\def\bar#1{\overline{#1}}

\def\inv{^{\raise.15ex\hbox{${\scriptscriptstyle -}$}\kern-.05em 1}}
\def\lbar{{\lower.35ex\hbox{$\mathchar'26$}\mkern-10mu\lambda}} 

\def\to{\rightarrow}

\let\<=\langle
\let\>=\rangle

\let\+=\uparrow

\let\Ga=\Gamma

\let\De=\Delta

\let\th=\theta

\begin{document}
\thispagestyle{empty}
\vspace*{.5cm}
\noindent
\hspace*{\fill}{CERN-TH/2002-070}\\
\vspace*{2.5cm}

\begin{center}
{\Large\bf Proton Decay Signatures of Orbifold GUTs}
\\[2.5cm]
{\large Arthur Hebecker and John March-Russell}\\[.5cm]
{\it Theory Division, CERN, CH-1211 Geneva 23, Switzerland}
\\[.2cm]
(April 3, 2002)
\\[1.1cm]

{\bf Abstract}\end{center}
\noindent
In grand unified theories based on orbifold constructions in higher 
dimensions, Higgsino-mediated proton decay is absent. However, proton decay 
mediated by $X$ and $Y$ gauge bosons is typically enhanced to levels 
detectable by current and future experiments. We analyse the phenomenology 
of proton decay induced by the minimal coupling of $X,Y$ gauge bosons. In 
particular, we show that the novel realization of matter in orbifold GUTs 
can lead to unusual final state flavour structure, for example, the dominance 
of the $p\to K^0\mu^+$ mode. Furthermore, we discuss proton decay induced by 
higher-derivative brane operators, finding potentially observable rates for 
natural values of the operator coefficients. 
\newpage

\setcounter{page}{1}

\section{Introduction}

Grand unified theories (GUTs) provide an elegant explanation of the origin 
of the three Standard Model (SM) gauge interactions and the fermion quantum
numbers~\cite{GG}. In their minimal supersymmetric extension, GUTs also lead 
to a remarkably successful prediction of $\sin^2\th_w$. This supports both 
the existence of supersymmetry (SUSY), broken near the weak scale, and some 
form of gauge-coupling unification at a scale $M_{\rm GUT}
\simeq 10^{16}\gev$. The 
idea of GUTs can also naturally accommodate relations among Yukawa couplings, 
leading most notably to the quite successful prediction of $m_b/m_\tau$.

Among the less attractive features of GUTs are the complicated Higgs sector 
required for realistic breaking of the gauge group and the necessity of 
modifying the unsuccessful first and second generation analogues of the
$m_b/m_\tau$ mass predictions. In addition, dimension-5 proton decay
operators arising from the exchange of supermassive coloured Higgsinos
severely constrain minimal SUSY GUT models (see e.g.~\cite{hitoshi}). 

Recently an elegant solution to these problems has been 
proposed in the context of SU(5)~\cite{kaw,AF,HN,HMR,HMR2} and 
SO(10)~\cite{abc,hnos} unification. The GUT gauge symmetry is now
realized in 5 or more space-time dimensions and broken 
to the SM group by compactification on an orbifold, utilizing boundary 
conditions that violate the GUT-symmetry. In the most studied case of
5 dimensions both the GUT group and 5d supersymmetry are broken
by compactification on $S^1/(Z_2\times Z_2')$, leading to an N=1 SUSY
model with SM gauge group. This construction allows one to avoid some 
unsatisfactory features of conventional GUTs with Higgs breaking, such as 
doublet-triplet splitting, dimension-5 proton decay, and Yukawa 
unification in the first two generations, while maintaining, at least at 
leading order, the desired gauge coupling unification~\cite{HN,HMR,CPRT}.

Higgsino mediated proton decay is absent because the triplets acquire mass 
via the KK expansion of 5d terms of the form 
\beq
{\cal L}\supset\int d^2\th \left(H_3^c \nabla_5 H_3 + H_{\bar{3}}^c\nabla_5
H_{\bar{3}}\right) + {\rm h.c.}
\label{eq:tripletmass}
\eeq
Thus, the mass couples the triplet Higgsino to a state in $H_3^c$, and not 
in $H_{\bar{3}}$.  Unlike the $H_3$ and $H_{\bar{3}}$ states, the $H_3^c$ 
and $H_{\bar{3}}^c$ fields do not couple directly to the quark and lepton 
superfields and the dangerous dimension-5 operators are absent. This 
absence can be viewed as a consequence of a U(1)$_R$ symmetry of the 5d 
theory.  An extension of the U(1)$_R$ to matter fields localized on the 
branes leads to $R$-parity, prohibiting baryon number violation at 
dimension 4 as well~\cite{HN}.

Although there is no dimension-4 or -5 proton decay, we argue in this 
letter that $X,Y$ gauge-boson mediated proton decay is now much more 
interesting.  First, the mass scale $M_c=1/R$ of the $X$ and $Y$ gauge
bosons in the orbifold GUT theories is lower than in conventional SUSY
GUTs, possibly approaching $10^{14}\gev$.\footnote{The reason for this
is that the running of differences of gauge couplings does not stop at
$M_c$, but continues in only a slightly modified way~\cite{HN,HMR}. 
So $M_c$ is lower than the unification scale.} 
This enhances the dimension-6 
proton decay processes to a level that may be seen in current and future 
experiments.  Second, the novel realization of the matter 
multiplets in orbifold GUTs changes in a striking way the signatures of 
$X,Y$ mediated proton decay and leads to a clear experimental 
distinction between orbifold and 4d GUT predictions.\footnote{We thank
Y.~Nomura for bringing to our attention Ref.~\cite{Nomura} where these
interesting features of orbifold GUT theories were previously noted.
In Sect.~3 we will comment on the relations of our respective
findings.}

We begin in Sect.~2 by deriving the dimension-6 proton decay operators 
arising from the minimal couplings of $X,Y$ gauge bosons.  We integrate
out the 5d $X,Y$ states in a 4d superfield formalism. The 
advantage of this approach is that no ill-defined contact-interaction 
$\sim\delta(0)$ (cf.~\cite{mp}) appears in the intermediate stages of the 
calculation.  Using this formalism we then discuss the effect of
higher-derivative brane operators involving the $X,Y$ gauge fields.
Such interactions, which are not forbidden by the gauge 
symmetries of the model, can lead to sizeable proton decay rates 
even in scenarios which are otherwise safe due to the location of the 
matter multiplets.  In Sect.~3, we calculate the corresponding
proton decay rates for the two classes of operators.  The obtained
results depend crucially on the location of matter 
multiplets on the two branes or in the bulk. In particular, we point out 
that in one of the most attractive realizations of the SU(5) orbifold
model, the minimal $X,Y$ couplings favour $p\to\mu^+K^0$ decay.  This
represents a striking orbifold GUT signature.  We also show
that, depending on the realization of matter, the higher-derivative
operators involving $X,Y$ gauge bosons, can be the numerically dominant source
of proton decay.  Overall, $X,Y$-mediated proton decay can occur at rates
observable in the current or next generation of proton decay experiments. 
Our conclusions are given in Sect.~4.

\section{Integrating out the $X,Y$ gauge bosons}

We begin by recalling the basic structure of the Kawamura
model~\cite{kaw}, which is based on a 5d super Yang-Mills theory on 
$I\!\!R^4\times S^1$, where the $S^1$ is 
parameterized by $y\in[0,2\pi R)$. The field space is then restricted by 
imposing the two discrete $Z_2$ symmetries, $y\to -y$ and $y'\to -y'$ (with 
$y'=y-\pi R/2$). The action of the $Z_2$'s in field space is specified 
by the two gauge twists $P$ and $P'$. If the original gauge group is SU(5) 
and the gauge twists are chosen as $P=1$ and $P'=$ diag$(1,1,1,-1,-1)$, 
an effective low energy theory with SM gauge symmetry results.

The supersymmetric version of this model can be described in terms of 4d 
superfields~\cite{mss}. A manifestly gauge-invariant form of the non-abelian 
action is given by~\cite{heb}
\beq
S=\int d^4x\int_0^l dy\frac{1}{2g_5^2}\mbox{tr}\left\{\int_{\theta^2}W^2+
\mbox{h.c.}+\int_{\theta^2\bar{\theta}^2}\left(e^{-2V}\nabla_5e^{2V}\right)^2
\right\},
\label{act}
\eeq
where $V$ is a real superfield depending on the additional parameter $y=x^5$,
$W$ is the corresponding field strength superfield, $\nabla_5=\partial_5+ 
\Phi$ is the covariant derivative in $x^5$ direction, and $\Phi$ is an 
$x^5$-dependent chiral superfield. 

We will be interested in baryon-number-violating processes mediated by 
$X,Y$ gauge bosons in scenarios where SM fermions are localized on the SU(5) 
brane at $y=0$, on the SM brane at $y=l=\pi R/2$, or in the bulk. 
We begin with the simplest case of SU(5)-brane localized matter.
The coupling to a chiral superfield $\Psi$ at $y=0$ is described by
the lagrangian
\beq
{\cal L}_\Psi=\delta(y)\int_{\theta^2\bar{\theta}^2}\bar{\Psi}e^{2V}\Psi=
\delta(y)\int_{\theta^2\bar{\theta}^2}\left\{\bar{\Psi}\Psi+2V^AJ^A+
{\cal O}(V^2)\right\}\,,
\label{lps}
\eeq
where $J^A=$tr$(\bar{\Psi}T^A\Psi)$ is the matter current. Note that the 
SU(5) generators $\{T^A\}$ fall into the two subsets $\{T^a\}$ and $\{T^{
\hat{a}}\}$, the first one being the SM generators and the second the basis 
of the orthogonal complement. 

Given a current $J$ (corresponding to an external field $\Psi$), the 
equation of motion for $V$ is obtained by varying the action of 
Eqs.~(\ref{act}) and (\ref{lps}). In the $\Phi=0$ gauge and to leading 
order in $V$ it reads (our superspace conventions are those of~\cite{wb})
\beq
-\frac{1}{8}D_\alpha\bar{D}^2D^\alpha V-\partial_5^2V+g_5^2J\delta(y)=0\,.
\label{eom}
\eeq
If $\Psi$ is on-shell, $D^2\Psi=0$, one finds $D^2J=0$. 
It is self-consistent to assume that, in the 
presence of an $x^\mu$-independent current ($\mu=0,..,3$), we have an 
$x^\mu$ independent solution $V$ with a superspace dependence given by 
$V\sim J$. Then the first term in Eq.~(\ref{eom}) vanishes and we need to 
solve 
\beq
-\partial_5^2V^{\hat{a}}+g_5^2J^{\hat{a}}\delta(y)=0
\label{eomh}
\eeq
for the $V$ components corresponding to $X,Y$ gauge bosons. Given the 
boundary conditions $\partial_5V^{\hat{a}}(y=0)=V^{\hat{a}}(y=l)=0$, the 
solution is $V^{\hat{a}}=-g_5^2(l-y)J^{\hat{a}}$ for $y>0$. The 
function is continuous at $y=0$ but has a discontinuous first derivative. 
Inserting this into the original lagrangian, the following effective 
operator in the low-energy 4d effective theory is generated:
\beq
{\cal O}_1=-lg_5^2\int_{\theta^2\bar{\theta}^2}\sum_{\hat{a}}
(J^{\hat{a}})^2\,.
\eeq
We have checked explicitly that the four-scalar interaction contained in the 
superfield operator ${\cal O}_1$ agrees with the result of~\cite{mp}, where 
the calculation was done by first integrating out auxiliary fields and then 
summing the exchanged KK modes in the component formalism. The advantage of 
our superfield approach is that no ill-defined contact-interaction 
$\sim\delta(0)$ appears in the intermediate stages of the 
calculation\footnote{
This observation has also been made in the context of a 3d superfield 
description of 4d SUSY theories with boundary~\cite{erd}.
}. 
This could also have been achieved by summing KK modes and making use of the 
cancellations between physical and auxiliary field contributions enforced 
by the unbroken 4d SUSY at each KK level.

The four-fermion interaction contained in ${\cal O}_1$ reads
\beq
{\cal O}_1\supset -\frac{\pi^2}{4}\,\frac{g_4^2}{M_c^2}\,(\bar{\psi}_i
\bar{\psi}_j)(\psi_k\psi_l)\sum_{\hat{a}}T^{\hat{a}}_{ik}T^{\hat{a}}_{jl}\,,
\label{fourfermi}
\eeq
where $\psi$ is the fermion from $\Psi$, $g_4^2=g_5^2/l$ is the 4d gauge 
coupling, and $M_c=1/R$ is the compactification scale which, at the same 
time, is the mass of the lowest-lying KK mode of $X,Y$ gauge bosons. Note 
that, up to the prefactor $\pi^2/4$, this is precisely the result expected
in a 4d GUT with gauge boson mass $M_c$. The prefactor can be easily 
understood as $2\,(\pi^2/8)$, where the 2 comes from the normalization of 
the KK modes due to their non-trivial bulk profile~\cite{HN} and 
$\pi^2/8=\sum_{n=1}^{\infty}(2n-1)^{-2}$ accounts for exchanging the full KK 
tower rather than just the lowest-lying mode.

We now turn to the case where SM fermions are localized on the SM
brane~\cite{HMR}.\footnote{Following~\cite{HMR} such realizations of
matter have also been used in recent TeV scale SU(3) models~\cite{su3}.}
Although in this case, $X,Y$ mediated proton decay is no
longer a generic prediction of the theory, it will occur if brane
operators of the type
\beq
{\cal
L}_\Psi=\frac{c}{M}\delta(y')\int_{\theta^2\bar{\theta}^2}\bar{\Psi}_1
\left(\nabla_5 e^{2V}\right)\Psi_2+\mbox{h.c.}\label{bbc}
\eeq
are included in the action~\cite{heb}. Here $\Psi_1$ and $\Psi_2$ are SM
superfields and the multiplication with $(\nabla_5 e^{2V})$ is defined by
using their standard embedding into SU(5) multiplets. Note that, even
though the $X,Y$ components of $V$ vanish at the SM brane, their $\partial_5$
derivatives appearing in $(\nabla_5 e^{2V})$ are non-zero. The prefactor
includes a dimensionless ${\cal O}(1)$ coefficient $c$ and the fundamental
scale $M\gg M_c$, which is required for dimensional reasons. Parametrically,
$M$ can not be larger than $\sim 1/g_5^2$, since this is the scale at which
the 5d gauge theory becomes strongly coupled.

Following the same line of reasoning as in the case of SU(5)-brane fermions,
one arrives at the analogue of Eq.~(\ref{eomh}), which is
\beq
-\partial_5^2V^{\hat{a}}-\frac{cg_5^2}{M}J^{\hat{a}}\partial_5\delta(y)=0\,,
\eeq
with the current $J^{\hat{a}}=$tr$(\bar{\Psi}_1T^{\hat{a}}\Psi_2+$h.c.). The 
solution is given by
\beq
V^{\hat{a}}(y)=\frac{cg_5^2}{M}J^{\hat{a}}\int_y^ld\tilde{y}\,
\delta(\tilde{y})\,.
\eeq
Inserting this into the original lagrangian, the following effective 
operator in the low-energy 4d effective theory is obtained:
\beq
{\cal O}_2=-\frac{c^2g_5^2}{M^2}\int_{\theta^2\bar{\theta}^2}\sum_{\hat{a}}
(J^{\hat{a}})^2\,\int_0^ldy\delta(y)^2=-\frac{\pi b c^2g_4^2}{2MM_c}
\int_{\theta^2\bar{\theta}^2}\sum_{\hat{a}}(J^{\hat{a}})^2\,.
\label{deriv}
\eeq
Here, to obtain the final expression, we have assumed the $\delta$ function 
to be localized on a scale $1/M$ and, accordingly, replaced the $y$ integral 
by $b M$, with $b$ an ${\cal O}(1)$ coefficient depending on the 
specific way in which the $\delta$ function is regularized. Thus, the 
operator ${\cal O}_2$ depends via $c^2$ on the a priori unknown strength 
of the brane-bulk coupling, Eq.~(\ref{bbc}), and via $b$ on the UV 
completion of the theory. This UV sensitivity can also be understood by 
observing that the derivative coupling of Eq.~(\ref{bbc}) enhances the 
contribution of higher KK modes so that the sum diverges. Parametrically, 
${\cal O}_2$ is suppressed with respect to ${\cal O}_1$ by a factor $M/M_c$, 
which can be as small as $\sim 10$ (allowing at least some range of validity 
for the effective 5d theory) or as large as $\sim 10^3$ (extending the 5d 
theory all the way to its strong coupling scale). Nevertheless,
as we argue in the next section, in models 
where the first generation is not on the SU(5) brane, ${\cal O}_2$ can be 
a competitive or even the dominant source of proton decay. 

In models where both branes contain SM fermions, the combined effect of the 
couplings of Eqs.~(\ref{lps}) and (\ref{bbc}) will give rise to further 
proton decay operators involving both SU(5)- and SM-brane fermions. The 
calculation is a straightforward combination of the two basic cases 
discussed above.

\section{Proton decay rates}

In SU(5) orbifold GUTs, there are three possible locations for matter:
the SU(5) brane, the bulk (which is also SU(5) symmetric), and the SM brane.
The various possible models are characterized by the placement of the $T_i$ 
($\bf 10$'s) and ${\bar F}_i$ ($\bf\bar{5}$'s) that make up the three 
($i=1,2,3$) generations of quarks and leptons. To reproduce the successful 
SU(5) $m_b/m_\tau$ mass relation, $T_3$ and ${\bar F}_3$ must reside on the
SU(5) brane. Furthermore, due to the different normalization of 4d and 5d
fields, Yukawa interactions involving one or two bulk matter fields are 
effectively suppressed by factors $(M/M_c)^{1/2}$ and $(M/M_c)$~\cite{HMROS} 
(recall that $M$ is the fundamental UV scale of the theory). Thus, both the 
relative lightness of the 2nd and 1st generations and the failure of the 
SU(5) mass predictions indicate that for $i=1,2$ either $T_i$ or $\bar{F}_i$ 
or both should be in the bulk or on the SM brane.

First we consider the case of proton decay operators arising from minimal
$X,Y$ gauge boson interactions.

Both for matter on the SM brane and in the SU(5)-symmetric bulk, minimally 
coupled $X,Y$ gauge bosons do {\em not} lead to dimension-6 baryon number 
violating operators~\cite{Nomura,Csaki}.  For matter on the
SM brane this is simply because the 
orbifold boundary conditions imply the vanishing of the $X,Y$ gauge boson 
wavefunctions on the SM brane. For bulk matter the situation is slightly more 
involved. Although 5d fields come in complete SU(5) multiplets, the orbifold 
projections $P$ and $P'$ imply that the zero modes under this action do not 
fill out full SU(5) multiplets.  From a ${\bf 10}$ we just get ${\bar U}$ 
and ${\bar E}$ zero modes, while from a ${\bf\bar{5}}$ we get $\bar{D}$. The 
remaining components of a full SM generation are realized at the zero mode 
level by taking another copy of ${\bf 10}$ and ${\bf \bar{5}}$ in the bulk 
and flipping the overall sign of the action of $P'$ on these multiplets. 
As a result, we have zero modes which fill out the full matter content of 
$T$ or $\bar{F}$ at the zero mode level~\cite{HN,HMR}. However, since 
the components of $T$ or $\bar{F}$ arise from different 5d SU(5) parent 
multiplets, the interaction of $X$ and $Y$ gauge bosons with these `split' 
multiplets do not convert SM quarks into SM 
leptons.
Rather, they convert SM fermions into superheavy ($m\sim 1/R$) 
exotic partner states.  Only with multiple $X,Y$ exchange can a 
baryon-number-violating operator involving only SM fields be produced.
Such operators are at minimum of dimension 8 and are irrelevant for 
proton decay experiments in the foreseeable future.\footnote{If
some or all of the SM fermions come from split multiplets localized in 
the bulk, $X,Y$-boson-mediated proton decay via dimension-6 
operators will occur if couplings analogous to Eqs.~(\ref{lps}) and 
(\ref{bbc}) mixing different bulk SU(5) multiplets are present. One can 
think of such terms as non-diagonal brane-localized kinetic terms (in a 
basis chosen to make the bulk kinetic term diagonal).  We do not consider
this case in detail here.} Thus, we now focus 
on proton decay due to interactions of $X,Y$ gauge bosons with SU(5)-brane 
matter.

The only $\De B= 1$ operators induced by single $X$ or $Y$ gauge boson 
exchange are 
\bea
{\cal O}_{TF} &=& \frac{\pi^2}{4}\,\frac{g_4^2}{M_c^2}\,\sum_{i,j}
a_i b_j \,\,\bar{d_R}_i \bar{u_R}_j \,\, L_{Li} Q_{Lj} \\
{\cal O}_{TT} &=& \frac{\pi^2}{4}\,\frac{g_4^2}{M_c^2}\,\sum_{i,j}
b_i b_j \,\,\bar{e_R}_i \bar{u_R}_j \,\,Q_{Li} Q_{Lj}\,.
\eea
Here and below our Weyl spinor notation is such that $\psi_{\rm Dirac}=(
\psi_L,\bar{\psi}_R)^T$. The operators ${\cal O}_{TF}$ and ${\cal O}_{TT}$
arise from gauge exchange between $T$ and $\bar{F}$ and between two $T$
multiplets respectively. The generation indices $i,j$ label the matter 
fields in the `locality basis,' \cite{HMROS} (in which the fields are 
defined by their location) so that $a_i=1$ or $a_i=0$ depending on whether 
$\bar{F}_i$ is on the SU(5) brane or not. Similarly $b_i=1,0$ depending on 
whether $T_i$ is localized on the SU(5) brane, or not. 

The fields in the locality basis (which is also the gauge basis) get mass 
via Yukawa interactions characterized by the matrices $\lambda_u$, 
$\lambda_d$, and $\lambda_e$. Low-energy experiments determine only their 
diagonal form, $\lambda_u^{\rm diag}$, $\lambda_d^{\rm diag}$, 
$\lambda_e^{\rm diag}$, and the CKM matrix, $V_{\rm CKM}$: 
\beq
\lambda_u^{\rm diag}=L_u^\dagger \lambda_u R_u , \qquad
\lambda_d^{\rm diag}=L_d^\dagger \lambda_d R_d , \qquad
\lambda_e^{\rm diag}=L_e^\dagger \lambda_e R_e , \qquad
V_{\rm CKM}=L_u^\dagger L_d .
\label{relations}
\eeq
The diagonalizing unitary rotations of type $L$ and $R$ act in generation 
space and connect the gauge and mass eigenstate bases. Denoting the latter 
by primes, the above $\De B=1$ operators take the form
\beq
{\cal O}_{TF} = \frac{\pi^2 g_4^2}{4M_c^2}\sum_{i,j} a_i b_j \left(
(\bar{d'_R}R_d^\dagger)_i (\bar{u'_R}R_u^\dagger)_j (\nu'_L)_i(L_d d'_L)_j
-(\bar{d'_R}R_d^\dagger)_i (\bar{u'_R}R_u^\dagger)_j (L_e e'_L)_i(L_u u'_L)_j
\right)
\label{op1}
\eeq
(note that, for our purposes, neutrinos can be treated as massless) and
\beq
{\cal O}_{TT} = \frac{\pi^2 g_4^2}{2 M_c^2}\sum_{i,j} b_i b_j 
(\bar{u'_R}R_u^\dagger)_j (\bar{e'_R}R_e^\dagger)_i
(L_u u'_L)_j (L_d d'_L)_i \, .
\label{op2}
\eeq
Here, in contrast to the minimal version of 4d GUTs, the coefficients of the 
operators involve combinations of elements of $L$ and $R$ matrices beyond 
those determined by $V_{\rm CKM}$. 

We start our discussion with the case considered originally in~\cite{kaw}: 
all generations located at the SU(5) brane. We use the approximation 
$L\simeq R\simeq \bf 1$ and focus on the mode $p\to\pi^0e^+$. The relevant 
operator reads
\beq
{\cal O}=\frac{\pi^2g_4^2}{4M_c^2}\left(2\,\bar{u_R'}\bar{e_R'}\,\,u_L'd_L'
-\bar{d_R'}\bar{u_R'}\,\,e_L'u_L'\right)\,,
\eeq
with only first generation fields. This gives rise to the decay rate
(see, e.g.,~\cite{his})
\beq
\Gamma(p\to\pi^0e^+)=\left(\frac{\pi^2}{4}\right)^2\frac{5\alpha^2m_p}{64\pi 
f_\pi^2}(1+D+F)^2\left(\frac{g_4^2A_R}{M_c^2}\right)^2\,.
\eeq
Taking the hadronic parameter $\alpha=0.015$ GeV$^3$~\cite{aok}, the pion 
decay constant $f_\pi=0.13$ GeV, the chiral perturbation theory parameters 
$D=0.80$ and $F=0.47$, the unified gauge coupling $g_4^2/(4\pi)=1/25$, and 
the renormalization coefficient $A_R=2.5$, the resulting life time is
\beq
1/\Gamma(p\to\pi^0e^+)=1.4\times 10^{34} \mbox{years} \times \left(\frac{M_c}
{10^{16}\mbox{GeV}}\right)^4\,.
\eeq
Using the Super-Kamiokande limit of $5.3\times 10^{33}$ years~\cite{shi,suz}, 
this leads to the bound $M_c\geq 0.8\times 10^{16}$ GeV~\cite{HN} (see 
also~\cite{dm}). Given the usual unifications scale $M_{\rm GUT}\simeq 
2\times 10^{16}$ GeV, almost no validity range for the higher-dimensional 
field theory is left. 

Let us now turn to a better motivated case where only the third generation 
is located on the SU(5) brane. This scenario predicts the large mass of the 
third generation, $b-\tau$ unification, and the absence of a similar mass 
relation in the first two generations. Given the observed structure of 
$V_{CKM}$, we assume that all matrices $L$ and $R$ are near the unit matrix. 
Thus, proton decay will be suppressed by the small off-diagonal elements of 
these matrices. To minimize the number of such elements, is advantageous to 
have an anti-neutrino and a strange quark in the final state. The relevant 
operator reads
\beq
{\cal O}=\frac{\pi^2g_4^2}{4M_c^2}\,\,\bar{s_R'}\bar{u_R'}\,\,\nu_{\tau L}'
d_L'\,\,(R_d^\dagger)_{23}(R_u^\dagger)_{13}(L_d)_{31}\,.
\eeq
The resulting decay rate, calculated using the analysis of~\cite{aok} and
including the phase space suppression due to the mass of the $K^+$, is
\beq
\Gamma(p\to K^+\bar{\nu}_\tau)=\left(\frac{\pi^2}{4}\right)^2\frac{\alpha^2
m_p}{72\pi f_\pi^2}\left(D\frac{m_p}{m_\Lambda}\right)^2\left(\frac{g_4^2A_R}
{M_c^2}\right)^2\left(1-\frac{m_{K}^2}{m_p^2}\right)^2
\left|(R_d^\dagger)_{23}(R_u^\dagger)_{13}(L_d)_{31}\right|^2\,,
\eeq
with $m_\Lambda=1.1$ GeV and $m_{K}=0.5$ GeV.  The parametric supression
by small mixing angles agrees with the estimate of \cite{Nomura}.
If, guided by $V_{CKM}$, 
one estimates the 2-3 and 1-3 mixing elements by 0.05 and 0.01 
respectively, one finds a life time 
\beq
1/\Gamma(p\to K^+\bar{\nu}_\tau)\simeq 6.6\times 10^{38} \mbox{years} \times 
\left(\frac{M_c}{10^{14}\mbox{GeV}}\right)^4\,.
\eeq
Thus, to observe an effect, one requires either experimental progress beyond 
the currently considered multi-megaton detectors (see e.g.~\cite{suz}),
or a very low compactification scale, or off-diagonal elements of the
unmeasured mixing matrices larger than the naive CKM-based estimate.

Finally, we consider what is arguably the best-motivated scenario, namely, 
a model with only $T_3$, $\bar{F}_3$ and $\bar{F}_2$ on the SU(5) brane. 
Following the discussion of~\cite{HMROS}, one finds that this model 
provides an explanation of all fermion mass hierarchies except for the 
lightness of the first-generation up-quark and electron and
the neutrinos in terms of the 
single parameter $(M/M_c)^{1/2}$. 
The dominant decay modes are $p\to K^0\mu^+$ and $p\to K^+\nu_\mu$.
For the striking $K^0\mu^+$ mode the relevant operator is
\beq
{\cal O}=-\frac{\pi^2g_4^2}{4M_c^2}\,\,\bar{s_R'}\bar{u_R'}\,\,\mu_L'u_L'\,\,
(R_u^\dagger)_{13}(L_u)_{31}\, . 
\eeq
Similarly we find that the $K^+\nu_\mu$ mode is also supressed by two 1-3
mixings.\footnote{In Ref.~\cite{Nomura} the modes $p\to K^+\nu_\mu$ and
$p\to \mu^+\pi^0$ were stated to be dominant with amplitudes supressed by
one 2-3 and two 1-3 mixings and by one 1-2 and two 1-3 mixings respectively.}
This gives rise to the rate
\beq
\Gamma(p\to K^0\mu^+)=\left(\frac{\pi^2}{4}\right)^2\frac{\alpha^2m_p}{32\pi 
f_\pi^2}\left(1+(D\!-\!F)\frac{m_p}{m_\Lambda}\right)^2\left(\frac{g_4^2A_R}
{M_c^2}\right)^2\left(1\!-\!\frac{m_{K}^2}{m_p^2}\right)^2\left|(
R_u^\dagger)_{13}(L_u)_{31}\right|^2\!\!.
\eeq
 The life time is
\beq
1/\Gamma(p\to K^0\mu^+)\simeq 2.1\times 10^{35} \mbox{years} \times \left(
\frac{M_c}{10^{14}\mbox{GeV}}\right)^4\,.
\eeq
Thus, in a phenomenologically well-motivated realization of this scenario,
with $M\simeq 10^{17}$ GeV and $(M/M_c)^{1/2}\simeq 30$, the next generation 
of proton decay experiments could `discover' orbifold GUTs via the striking 
signal of a $K^0\mu^+$ final state.  
(Note that in the case where $T_2$ but not ${\bar F}_2$ is on the SU(5)
brane proton decay could also be discovered by the striking $p\to K^0 \mu^+$
mode. In this case the mixing supression is only by two
1-2 mixings~\cite{Nomura}.)

We now turn to the non-minimal case where proton decay arises from the
higher-derivative interactions of the $X,Y$ bosons with matter on the 
SM brane.  This is of interest especially in the minimal GUT models
where all matter in localized on the SM brane~\cite{HMR}.  The proton
decay rate that results from the operator Eq.(\ref{deriv}) is
\beq
1/\Gamma(p\to\pi^0e^+)=3.5\times 10^{34} \mbox{years} \times \frac{1}{b^2 c^4}
\left(\frac{M_c}{10^{15}\mbox{GeV}}\right)^2
\left(\frac{M} {10^{17}\mbox{GeV}}\right)^2\,.
\eeq
Assuming natural ${\cal O}(1)$ values for the unknown coefficients
$b$ and $c$, this leads to a potentially observable rate.  This same mechanism
can also apply to split multiplets, but in this case there is a further
suppression of $M_c/M$ in the rate for each bulk field.

\section{Conclusions}

Proton decay is one of the most characteristic and striking signatures
of conventional 4d GUTs.  In orbifold GUTs, although dimension 5 operators
are naturally absent, proton decay induced by the exchange of $X,Y$
boson states is enhanced.  Moreover, the minimal gauge interactions of
$X,Y$ gauge bosons only induce $\De B=\De L=1$ operators for those matter
multiplets localized to the SU(5)-invariant brane.  For example, in
models with the most realistic flavour
structure, the 1st generation is not localized
to the SU(5) brane, and proton decay arises from the mixing of
brane and bulk matter to form mass eigenstates.  As a result,
minimal $X,Y$ gauge interactions favour unusual final states, e.g.
$p\to K^0\mu^+$, at a potentially observable rate.  Thus the rates and
flavour structure of proton decay processes provide important probes of
orbifold GUTs.
In addition to minimal $X,Y$ couplings, there can exist
higher-derivative brane-localized couplings
of $X,Y$ gauge bosons directly to 1st generation states.  Although
formally suppressed by $M_c/M$, these new operators do not rely on
mixing and can be the numerically dominant source of proton decay, with rates
$\Ga(p\to \pi^0 e^+)$ observable in current and future experiments.

\end{document}